\documentclass[twocolumn,prd,superscriptaddress,showpacs]{revtex4}
\usepackage{graphicx}
\usepackage{dcolumn}
\usepackage{bm}
\begin{document}
\title{Factorization fits to charmless strangeless $B$ decays}
\author{W. N. Cottingham}
\affiliation{H. H. Wills Physics Laboratory, Bristol University,
Tyndall Avenue, Bristol BS8 1TL, United Kingdom}
\author{I. B. Whittingham}
\affiliation{School of Mathematical and Physical Sciences,
James Cook University, Townsville, Australia, 4811}
\affiliation{H. H. Wills Physics Laboratory, Bristol University,
Tyndall Avenue, Bristol BS8 1TL, United Kingdom}
\author{F. F. Wilson}
\affiliation{Rutherford Appleton Laboratory, Chilton, Didcot,
OX11 0QX, United Kingdom}
\date{\today}
\begin{abstract}
We present fits to charmless strangeless hadronic $B$ decay data
for mean branching ratios and $CP$-violating asymmetries
using the QCD factorization model of Beneke \textit{et al.}
Apart from one $CP$-violating parameter, the model gives a very good
representation of 26 measured data. We find the CKM angle
$\alpha =(93.5\pm 8.4 - 1.3)^{\circ}$ and to be quite stable to
plausible "charming penguin" corrections.
\end{abstract}

\pacs{12.15.Ji, 12.39.St, 13.25.Jx}
\maketitle

\section{\label{sec:intro} Introduction}

A wealth of experimental data on hadronic charmless $B$ decays has become
available from the BaBar and Belle experiments.  Many new branching ratios
and $CP$-violating parameters have been measured within more precise
error limits.
These studies of the numerous $B$ decay channels are designed to test the
Cabbibo-Kobayashi-Maskawa (CKM) explanation of $CP$ violation in the
standard model.

We have made previous attempts \cite{Groot,Cott} to understand charmless $B$
meson decay data in terms of the QCD factorization model of Beneke
\textit{et al.} (BBNS) \cite{BBNS,BenekePV}.
In fitting the data we found evidence
for \textit{charming penguin-like contributions} \cite{charming} to the
decay amplitudes in addition to the BBNS amplitudes.
The charming penguin
contribution to the strangeless modes is suppressed relative to the strange
modes by one power of the Wolfenstein parameter $\lambda \approx 0.22$.
In \cite{Groot} we attempted a simultaneous fit to both strange and
non-strange channels and, although we obtained a satisfactory fit to
measured branching ratios and some $CP$-violating asymmetries, the predicted
CKM angles deviated significantly from other analyses.
The strange channels in isolation provide the best data for
examining the phenomenology of charming penguins \cite{Buras2000}. 
With this information their
influence on the strangeless modes should then be investigated.

We report here the results of our application of a BBNS analysis to charmless
strangeless $B$ decays.
We use the method and notation of \cite{Groot}
and attempt to fit the data on the mean branching ratios and
$CP$-violating asymmetries of the decays $B \rightarrow \pi \pi, \pi \rho ,
\pi \omega $ and $\rho \rho $, together with $\sin (2 \beta )$,
26 independent
measurements in all. We include many more $CP$ asymmetry measurements in
our data set and also three new $\rho \rho $ channels which exhibit the
expected helicity-zero dominance. The paper is organized as follows. In
Sec. II we briefly discuss the structure of the decay amplitude within
QCD factorization.
The weak annihilation contributions are summarized in
Sec. III and the method and results of our best fit are presented in Sec. IV.
Sec. V contains our discussion and conclusions.

\section{Decay amplitude in QCD factorization}

In QCD factorization, the matrix elements of the operators in the
effective Hamiltonian $\mathcal{H}_{\text{eff}}$ are separated into
short distance contributions at scale $O(1/m_{b})$ that are
perturbatively calculable and long distance contributions
$O(1/\Lambda_{\text{QCD}})$ that are parametrized. The
amplitude for $B$ decay into two
light hadrons (mesons) $M_{1,2}$ has the form, neglecting weak
annihilation processes, \cite{BBNS,Cott,Groot}
\begin{eqnarray}
\label{decay-amp}
\langle M_{1}M_{2}|\mathcal{H}_{\text{eff}}|B \rangle & = &
 \frac{G_{F}}{\sqrt{2}}
\{[\sum_{i=1,2} \lambda_{u}a^{u}_{i}
+\sum_{p=u,c}\sum_{i=3,\ldots ,6,9} \lambda_{p}a^{p}_{i}] \nonumber  \\
& & \times [T_{i}(M_{1},M_{2}) + T_{i}(M_{2},M_{1})] \}
\end{eqnarray}
where $\lambda_{p}=V^{*}_{pd}V_{pb}$ is a product of CKM matrix elements.
In Eq. (\ref{decay-amp}) we have included current-current tree processes
represented by the Wilson coefficients $C_{1,2}$, QCD penguin processes
represented by $C_{3,\ldots ,6}$ and the dominant
electroweak penguin process represented by $C_{9}$.
The factorization matrix elements $T_{i}(M_{1},M_{2})$ involve
products of two-quark current matrix elements and, neglecting factors
$(m_{P,V}/m_{B})^{2}$ and assuming zero helicity dominance for
$B \rightarrow V_{1}V_{2}$ decays, have the form \cite{Cott}
\begin{equation}
\label{tfact}
T_{i}(M_{1},M_{2})=c\;m_{B}^{2}f_{M_{1}}F_{M_{2}}
\end{equation}
where $f_{M}$ are the well determined electroweak decay constants
$\{f_{\pi},f_{\omega},f_{\rho}\}$, $F_{M}$ are the $B$ transition form
factors $\{F_{\pi},A_{\omega },A_{\rho}\}$ and the constant $c$ is a product
of factors $1,\pm 1/\sqrt{2},$ e.t.c. from the flavor composition of
the $B$ and $M_{1,2}$ mesons.

The factorization coefficients $a_{i}$ are calculated from the
Wilson coefficients $C_{i}(\mu )$ at a scale $\mu $ of $O(m_{b})$
and have the form \cite{BBNS}
\begin{equation}
\label{acofs}
a_{i}(M_{1},M_{2})=a_{i,I}(M_{2})+a_{i,II}(M_{1},M_{2})
\end{equation}
where $M_{1}$ is the recoil meson containing the spectator (anti) quark
and $M_{2}$ is the emitted meson. The complex quantities $a_{i,I}$ describe
the formation of $M_{2}$, including nonfactorizable corrections from
hard gluon exchange or light quark loops in penguins. The hard gluon
exchanges with the spectator quark are described by the
(possibly) complex quantities $a_{i,II}$. In these correction terms
the leading-twist light cone distribution
functions for both
pseudoscalar and vector mesons are expanded in the first few terms of a
Gegenbauer expansion
\begin{equation}
\label{Gegen}
\Phi_{M}(x,\mu ) = 6x (1-x) [1+\sum_{n} \alpha^{M}_{n} C^{3/2}_{n}(2x-1)].
\end{equation}
The light cone corrections $\alpha^{M}_{n}$ are
anticipated to be small but they are not well established. Consequently it
is common to use the asymptotic form $\Phi_{M}(x,\mu )=6x(1-x)$, valid for
the mass scale $\mu \rightarrow \infty $, in applications of QCD
factorization. All coefficients $a_{i,I}$ except $a_{6,I}$ are then the
same for all decays.
If the corrections $\alpha^{M}_{1,2}$ are included, the coefficients
most affected are $a_{2,I}$ and $a_{4,I}$ (see Table IV of \cite{Groot}).

The contributions of the $a_{i,II}$ coefficients to the decay amplitudes
have the form
\begin{equation}
\label{aiII}
f_{M_{2}}F_{M_{1}}a_{i,II} = \frac{4 \pi }{9}\epsilon_{i}C_{i^{\prime}}
\alpha_{s} \beta_{i}
\end{equation}
where $i^{\prime}=i-(-1)^{i}$, $\epsilon_{i}=+1 (i=1, \ldots ,4,9),
\epsilon_{5}=-1, \epsilon_{6}=0$,  and
\begin{eqnarray}
\label{betai}
\beta_{i} & = & \frac{f_{B}f_{M_{2}}f_{M_{1}}}{m_{B}\lambda_{B}}
\left[3(1+\epsilon_{i}\alpha^{M_{2}}_{1}+\alpha^{M_{2}}_{2})
(1+\alpha^{M_{1}}_{1}+\alpha^{M_{1}}_{2}) \right. \nonumber  \\
&& + \left. r^{M_{1}}_{\chi }(1-\epsilon_{i}\alpha^{M_{2}}_{1}+
\alpha^{M_{2}}_{2})X^{M_{1}}_{H})\right] .
\end{eqnarray}
Here $r^{M_{1}}_{\chi}$ are chiral factors, $X^{P}_{H}=X_{H}$
and \cite{BenekePV}
\begin{equation}
\label{xvh}
X^{V}_{H}= 3(\alpha^{V}_{1}+\alpha^{V}_{2})X_{H} -
(6+9\alpha^{V}_{1}+11 \alpha^{V}_{2}).
\end{equation}
The non-perturbative complex parameter $X_{H}$
is the contribution of a logarithmic
end-point divergence in the integration over the twist-3 light cone
distribution function
\begin{equation}
\label{XH}
X_{H} = \int ^{1}_{0}\frac{dx}{1-x},
\end{equation}
and is usually parameterized as
\begin{equation}
\label{XH-par}
X_{H}=\ln \left( \frac{m_{B}}{\Lambda_{\text{QCD}}}\right)
+ \rho_{H}e^{i\phi_{H}}
\end{equation}
where $\ln (m_{B}/\Lambda_{\text{QCD}})=3.03$.
The coefficients $a_{i,II}$ are not universal even when light cone
corrections are neglected and they contain the parameter $X_{H}$
which is only loosely constrained by model estimations.
These $a_{i,II}$ contributions to the decay
amplitudes are independent of the $B$ transition form factors but do
involve the poorly determined parameter $f_{B}/\lambda_{B}$ where
$f_{B}$ is the $B$ leptonic decay constant and $\lambda_{B} \approx 0.35$
GeV is related to the $B$ light cone distribution function.
We note that substantial
light cone corrections $\alpha^{M}_{1,2}$ can significantly enhance
the $a_{i,II}$ coefficients.

The energies
involved in the calculation of $a_{i,II}$ imply that the appropriate
scale is not that of the scale $\mu $ used in calculating the $a_{i,I}$
but $\mu _{h} = \sqrt{\Lambda_{h} \mu }$ where \cite{BBNS}
$\Lambda_{h}=0.5$ GeV. For the choice $\mu =m_{b}/2$ this gives
$\alpha_{s}f_{B}/(m_{B}\lambda_{B})=0.0209$.

\section{Annihilation contributions}

$B$ meson decay can also be initiated by $b$ quark annihilation with its
partner. Although the annihilation contributions to the decay amplitude
are formally of $O(\Lambda_{\text{QCD}}/m_{b})$ and power suppressed, they
violate QCD factorization because of end point divergences. However these
weak annihilation contributions can be included into the decay amplitudes
by treating the end point divergences as phenomenological complex parameters
$X_{A}$.

The annihilation contribution to the decay amplitude is
\begin{eqnarray}
\label{annih}
\langle M_{1}M_{2}|\mathcal{H}^{\text{ann}}_{\text{eff}}|B \rangle & = &
\frac{G_{F}}{\sqrt{2}}B_{M_{1}M_{2}} \{\lambda_{u}
[d_{1}C_{1}+d_{2}C_{2}]A^{i}_{1} \nonumber  \\*
&&-\lambda_{t}\{d_{3}[C_{3}A^{i}_{1}+C_{5}A^{i}_{3}]  \nonumber  \\*
&&+[d_{4}C_{4}+d_{6}C_{6}]A^{i}_{1} \nonumber  \\*
&&+d_{5}[C_{5}+N_{c}C_{6}]A^{f}_{3}\}
\end{eqnarray}
where
\begin{equation}
\label{Bcoeff}
B_{M_{1}M_{2}}= \frac{C_{F}}{N_{c}^{2}}f_{B}f_{M_{1}}f_{M_{2}},
\end{equation}
$C_{F}=(N_{c}^{2}-1)/2N_{c}$ and $N_{c}$ is the number of colors.
The quantities
$A^{i,f}_{1,3}(M_{1},M_{2})$, where the superscript $i(f)$ denotes gluon
emission from initial (final) state quarks, evaluated using the
asymptotic form of Eq. (\ref{Gegen}) are given by \cite{BenekePV} 
\begin{eqnarray}
\label{A-cofs}
A^{i}_{1}(P_{1},P_{2})&=& 2 \pi \alpha_{s}
\left[9\left(X_{A}-4+\frac{\pi^{2}}{3}\right)+r^{P_{1}}_{\chi}
r^{P_{2}}_{\chi}X^{2}_{A}\right],  \nonumber  \\
A^{i}_{3}(P_{1},P_{2})&=& 6 \pi \alpha_{s}(r^{P_{1}}_{\chi}-
r^{P_{2}}_{\chi})(X^{2}_{A}-2X_{A}+\frac{\pi^{2}}{3}), \nonumber  \\
A^{f}_{3}(P_{1},P_{2}) &=& 6 \pi \alpha_{s}
(r^{P_{1}}_{\chi}+r^{P_{2}}_{\chi})(2X_{A}^{2}-X_{A}), \nonumber  \\
A^{i}_{1}(P,V)&=& 6 \pi \alpha_{s}
\left[3\left(X_{A}-4+\frac{\pi^{2}}{3}\right) \right. \nonumber  \\
& & + \left. r^{P}_{\chi} r^{V}_{\chi}(X_{A}^{2}-2X_{A})\right],  \nonumber  \\
A^{i}_{3}(P,V)& =& 6 \pi \alpha_{s}\left[ r^{P}_{\chi}\left(X_{A}^{2}-
2X_{A}+\frac{\pi^{2}}{3}\right) \right. \nonumber  \\
& & -3 \left. r^{V}_{\chi} \left(X_{A}^{2}-2X_{A}-
\frac{\pi ^{2}}{3}+4\right)\right],
\nonumber  \\
A^{f}_{3}(P,V)&=& 6 \pi \alpha_{s} [r^{P}_{\chi}(2X_{A}^{2}-X_{A}) \nonumber  \\
& & -3r^{V}_{\chi}(2X_{A}-1)(2-X_{A})],
\nonumber  \\
A^{i}_{1}(V_{1},V_{2}) &=& 6 \pi \alpha_{s}
\left[3\left(X_{A}-4+\frac{\pi^{2}}{3}\right)\right. \nonumber  \\
& & + \left. r^{V_{1}}_{\chi} r^{V_{2}}_{\chi}(X_{A}^{2}-2X_{A})\right],
\nonumber  \\
A^{f}_{3}(V_{1},V_{2})& =& 18 \pi \alpha_{s} (r^{V_{1}}_{\chi}+
r^{V_{2}}_{\chi})(2X_{A}-1)(X_{A}-2).
\end{eqnarray}
The coefficients $d_{i}(M_{1},M_{2})$ are Clebsch-Gordan type factors and
are  tabulated in \cite{Groot} for all the decays studied here apart from
$B \rightarrow \rho^{+}\rho^{-}$ for which the required $d$- coefficients
are $(1,0,-1,2,-1,2)$.

\section{Fitting method}

We attempt to fit the experimental data on the mean branching ratios and
$CP$-violating asymmetries of the decays $B \rightarrow \pi \pi, \pi \rho ,
\pi \omega $ and $\rho \rho $, together with $\sin (2 \beta )$.
This gives 26 independent measurements in all.
There are many parameters in the equations of the
BBNS theory that are not precisely known, we choose the $B$ semileptonic
transition form factors $F_{\pi},\;A_{\rho}$ and $A_{\omega}$, the
Wolfenstein CKM parameters $\rho $ and $\eta $ and the complex
non-perturbative hard scattering parameter $X_{H}$ and annihilation
parameter $X_{A}$ as our nine fitting parameters.

We assign to each of the 26 independent data a number $\alpha $ and
construct a $\chi ^{2}$ function of the nine fitting parameters
\begin{equation}
\label{chi2}
\chi^{2}  = \sum_{\alpha=1}^{26}[((\text{Theory})_{\alpha}-
(\text{mean data value})_{\alpha})/\sigma_{\alpha}]^{2}.
\end{equation}
$\sigma_{\alpha}$ is an experimental error formed by amalgamating
statistical and systematic errors. We then use a minimization procedure
based upon the program MRQMIN from \cite{Recipes}
to try to obtain acceptable fits of the theory to experiment. An
acceptable fit has a low value of $\chi^{2}$ at the minimum with
parameter values that lie within known acceptable limits.
Other parameters in the analysis such as the
correction factors $\alpha^{M}_{1,2}$ to the light cone
distribution functions of the participating mesons are
poorly known and  not easy to incorporate as variables in the
minimization procedure. In this paper we use three sets of values for
$\alpha^{M}_{1,2}$, the zero set and two sets suggested by theory
(those of \cite{BenekePV} and \cite{LuYang})
to examine the sensitivity of the
results to the choice of these correction factors. We also examine how the
results depend on our choice of the chiral factor $r^{\pi}_{\chi}$ and
the factor $\xi_{B} \equiv \alpha_{s} f_{B}/(m_{B}\lambda_{B})$
which enters into the
coefficients $a_{i,II}$ representing the spectator quark interactions.

Table \ref{fpars} shows the best fit values for the parameters,
together with the
statistical error from the $\chi^{2}$ and a systematic error from the
different light cone correction factors, e.t.c.  This fit
has a minimum  $\chi^{2}$ of 9.2 and is for the light
cone distribution functions of L\"{u} and Yang \cite{LuYang}, the chiral
factors $r^{\pi}_{\chi}=1.0$ and $r^{\rho}_{\chi}=0.1$, and
$\xi_{B} =0.0314$.
An unexpected feature of this
fit is the large value of the spectator quark interaction term
$\rho_{H}$ which was not expected to be greater than 3.0. It can be seen
however that its statistical error is large, the $\chi^{2}$ function
has a very shallow worm hole through parameter space. Constraining $\rho_{H}$
to be 3.0 gives a minimum $\chi^{2}$ of 12.5 and a quite acceptable fit. The
parameter changes this induces, the changes due to the different light
cone distribution functions, and those from taking $r^{\pi}_{\chi}=0.9$
or $\xi_{B}= 0.0209$, are summarized in the column of systematic errors.
These errors, like the statistical errors, are of course highly correlated
but they indicate the sensitivity of the parameters to different choices.
Our fit also implies that $V_{ub}=(3.89\pm 0.24)\times 10^{-3}$, to
be compared with $(3.7\pm 0.8 )\times 10^{-3}$.

The predicted mean branching ratios and $CP$ asymmetries for the parameter
values of Table \ref{fpars} are shown in Table \ref{fits}, together with
the individual contributions to the total minimum $\chi^{2}$ of 9.2.
Also shown are the estimated theoretical errors, the statistical errors    
are the $ 1 \sigma $ standard deviations from the $9 \times 9$ error
matrix of the $\chi^{2}$ minimization and the systematic errors arise
from changes in the light cone distribution functions, e.t.c.

If significant contributions, such as charming penguins, are needed to
understand the charmless strange $B$ decays then it is to be expected that
similar but smaller contributions should be added to the BBNS analysis. In
the spirit of \cite{Groot} we have investigated
the influence of charming penguins, as will be discussed in the next section.

\begingroup
\begin{table}
\caption{\label{fpars} Best fit values and errors for the $B$ transition
form factors $F_{\pi},\;A_{\rho}$ and $A_{\omega}$, the complex hard
scattering $X_{H}=3.03+\rho_{H}\exp(i\phi_{H})$ and annihilation
$X_{A}=x_{A}+iy_{A}$ parameters, and the CKM
angles $\alpha_{\text{CKM}} $ and $\beta_{\text{CKM}} $. }
\begin{ruledtabular}
\begin{tabular}{cddd}
Parameter &
\multicolumn{1}{c}{Mean} &
\multicolumn{1}{c}{Statistical} &
\multicolumn{1}{c}{Systematic} \\
 & \multicolumn{1}{c}{Value} &
\multicolumn{1}{c}{Error} &
\multicolumn{1}{c}{Error} \\
\hline
$F_{\pi}$ & 0.218 & \pm 0.022 & +0.015,-0.003\\
$A_{\rho}$ & 0.317 & \pm 0.027  &  -0.021\\
$A_{\omega }$ & 0.372 & \pm 0.050  & -0.03 \\
$\rho_{H}$ & 8.73 & \pm 2.82  &  +3.1,-5.26\\
$\phi_{H}$ & -2.05 & \pm 0.15  & +0.49,-0.16 \\
$x_{A}$ & -1.01 & \pm 0.50  & +0.02,-0.34\\
$y_{A}$ & 2.89 & \pm 0.85  & +0.73,-0.39 \\
$\alpha_{\text{CKM}}$ & 93.5^{\circ} & \pm 8.4^{\circ} & -1.3^{\circ} \\
$\beta_{\text{CKM}}$ & 24.2^{\circ} & \pm 2.3^{\circ} &
+0.2^{\circ},-0.4^{\circ}\\
\end{tabular}
\end{ruledtabular}
\end{table}
\endgroup

\begingroup
\begin{table*}
\caption{\label{fits} Measured branching ratios and $CP$ asymmetries
(Data), experimental error ($\sigma $), best fit theoretical values (Theory),
estimated theoretical errors (Statistical) and (Systematic),
contribution $\chi^{2}_{\alpha}$ to the
total minimum $\chi^{2}$ of 9.2, and the reference (Ref.) for the
experimental data.
All branching ratios are in units of $10^{-6}$.}
\begin{ruledtabular}
\begin{tabular}{cddddddl}
 & \multicolumn{1}{c}{Data} &
\multicolumn{1}{c}{$\sigma $} &
\multicolumn{1}{c}{Theory} & \multicolumn{1}{c}{Statistical} &
\multicolumn{1}{c}{Systematic} &
\multicolumn{1}{c}{$\chi^{2}_{\alpha}$} & Ref.  \\
&&&& \multicolumn{1}{c}{Error} & \multicolumn{1}{c}{Error} &&
\\
\hline
Br($\pi^{+}\pi^{-}$) & 4.6 & 0.4 & 4.73 & \pm 0.45 & +0.22,-0.07
& 0.11 & 8 \\
Br($\pi^{0}\pi^{0}$) & 1.2 & 0.4 & 1.15 & \pm 0.33 & +0.0,-0.38
& 0.02 & 9 \\
Br($\omega \pi^{0}$) & <1.0 & & 0.48 & \pm 0.30 & +0.02,-0.23
& 0.27 & 8 \\
Br($\pi^{\pm}\rho^{\mp}$) & 24.0 & 2.5 & 23.6 & \pm 1.54 & +0.20,-0.06
& 0.03 & 8 \\
Br($\pi^{0}\rho^{0}$) & 5.1 & 2.4 & 2.90 & \pm 0.44 & +0.0,-1.08
& 0.84 & 8 \\
Br($\rho^{+}\rho^{-}$) & 30.0  & 6.0  & 31.9 & \pm 4.02 & +0.0,-1.90
& 0.10 & 9 \\
Br($\rho^{0}\rho^{0}$) & <1 &  & 0.66  & \pm 0.23 & +0.0,-0.33
& 0.12 & 9 \\
Br($\pi^{\pm}\pi^{0}$) & 5.2  & 0.8  & 4.51 & \pm 0.65 & +0.71,-0.24
& 0.74 & 8 \\
Br($\omega \pi^{\pm}$) & 5.9 & 0.8 & 5.90 & \pm 1.00 & \pm 0.0
& 0.00 & 8 \\
Br($\pi^{\pm}\rho^{0}$) & 9.2  &  1.2  & 10.1 & \pm 1.03 & +0.40,-0.0
& 0.61 & 10 \\
Br($\pi^{0}\rho^{\pm}$) & 12.0  & 1.9  & 10.8 & \pm 0.80 & +0.30,-1.70
& 0.42 & 8 \\
Br($\rho^{0}\rho^{\pm}$) & 26.4 & 6.4  & 23.0 & \pm 2.19 & +0.0,-3.10
& 0.28 & 11 \\
$C_{\pi^{+}\pi^{-}}$ & -0.09  & 0.19  & -0.13 & \pm 0.03 & \pm 0.01
& 0.04 & 9 \\
$S_{\pi^{+}\pi^{-}}$ & -0.30  & 0.21  & -0.21 & \pm 0.14 & +0.0,-0.06
& 0.17 &  9 \\
$A_{\pi^{\pm}\pi^{0}}$ &  0.05  & 0.15  & -0.02 & \pm 0.0 & +0.01,-0.0
& 0.23 & 9  \\
$A_{\pi^{\pm}\rho^{0}}$  & -0.17  & 0.13  & -0.15  & \pm 0.08 & +0.0,-0.03
& 0.03 & 10 \\
$A_{\pi^{0}\rho^{\pm}}$  &  0.23  & 0.22  & 0.15 & \pm 0.06 & +0.03,-0.0
& 0.13 & 10 \\
$C_{\pi^{\mp}\rho^{\pm}}$  & 0.34  & 0.16  & 0.03 & \pm 0.01 & +0.0,-0.01
& 3.76 & 9  \\
$\Delta C_{\pi^{\mp}\rho^{\pm}}$  & 0.15  & 0.14  & 0.18 & \pm 0.08
& +0.0,-0.01 & 0.25 & 9 \\
$S_{\pi^{\mp}\rho^{\pm}} $  & -0.10  &  0.18  & -0.15 & \pm 0.15
& \pm 0.01 & 0.08 & 9 \\
$\Delta S_{\pi^{\mp}\rho^{\pm}}$  &  0.22  & 0.18  & 0.15 & \pm 0.05
& +0.02,-0.11 & 0.15 & 9 \\
$A_{\pi^{\mp} \rho^{\pm}}$  &  -0.09  &  0.06  &  -0.10 &\pm 0.03
& \pm 0.0 & 0.04 & 9 \\
$C_{\rho^{\mp} \rho^{\pm}}$  &  -0.23  &  0.38  & 0.03 & \pm 0.09
& \pm 0.0 & 0.47 & 9 \\
$S_{\rho^{\mp} \rho^{\pm}}$  &  -0.19  & 0.44 & -0.27 & \pm 0.02
& +0.04,-0.0 & 0.03 & 9 \\
$A_{\rho^{\mp} \rho^{0}}$  & -0.09  &  0.16  & -0.01 & \pm 0.0
& \pm 0.0 & 0.27 & 11 \\
$\sin (2 \beta )$  &  0.734  &  0.054 & 0.747 & \pm 0.046  & \pm 0.005 
& 0.06 & 8 \\
\end{tabular}
\end{ruledtabular}
\end{table*}
\endgroup

\section{Discussion and Conclusions}

Apart from the value of $\rho_{H}$, which has been discussed above, the
fitting parameters of Table \ref{fpars} are within their expected
ranges. In particular, the CKM parameters $V_{ub}$ and the angles
$\alpha_{\text{CKM}}$ and $\beta_{\text{CKM}}$ have very acceptable values
within quite tight bounds. All the experimental data, apart from
$C_{\pi^{\mp}\rho^{\pm}}$ are well fitted by the analysis. The largest
individual $\chi^{2}$ of 3.76 is for this $CP$-violating parameter
and it can be seen that our analysis gives a value much smaller than
the measured value, which is from the BaBar experiment.  The theoretical
result is in much better accord with the Belle measurement
$C_{\pi^{\mp }\rho^{\pm}}=0.25 \pm 0.17 ^{+0.02}_{-0.06} $
\cite{Beijing2004} (talk by M. Giorgi).
A remeasurement of this parameter would be of interest.

The angles $\alpha_{\text{CKM}}$ and $\beta_{\text{CKM}}$ are the most
important parameters to come out of the analysis.
It is of interest to compare our value of $\alpha_{\text{CKM}}$ with
those from isospin analyses and the Grossman-Quinn bound. These analyses
are independent of the details of the QCD penguin contributions. The
small $\rho^{0}\rho^{0} $ branching fraction makes the $\rho \rho $
analysis the most precise and yields $\alpha_{\text{CKM}} =96^{\circ}
\pm 10^{\circ}\pm 4^{\circ} \pm 11^{\circ}$ \cite{Beijing2004} (talk by
M. Giorgi) where the last error is from the Grossman-Quinn bound.
However our neglect in
particular of charming penguin contributions to the decay amplitudes
will bias our estimates. We have only made a preliminary assessment of
the influence of charming penguins, our conclusions so far are that the
small branching fractions like Br$(\pi^{0}\pi^{0})$ are somewhat
sensitive but the $CP$-violating data can be more affected, especially those
like $C_{\pi^{\mp}\rho^{\pm}}$ which our analysis predicts to be very small.
So far we have not found the angles $\alpha_{\text{CKM}}$ and
$\beta_{\text{CKM}}$ to be particularly sensitive to charming penguins
but more work is required to get reliable estimates of the systematic errors.

In conclusion, we find that the BBNS analysis so far does well in fitting
the data. We see no evidence in charmless strangeless decays for any
physics beyond the Standard Model and the analysis holds the promise of
giving tight constraints on the values of the CKM parameters.

\bibliographystyle{prsty}

\end{document}